\input harvmac

\def\Title#1#2{\rightline{#1}\ifx\answ\bigans\nopagenumbers\pageno0
\vskip0.5in
\else\pageno1\vskip.5in\fi \centerline{\titlefont #2}\vskip .3in}

\font\caps=cmcsc10

\noblackbox
\parskip=1.5mm

  
\def\npb#1#2#3{{\it Nucl. Phys.} {\bf B#1} (#2) #3 }
\def\plb#1#2#3{{\it Phys. Lett.} {\bf B#1} (#2) #3 }

\def\mpla#1#2#3{{\it Mod. Phys. Lett.} {\bf A#1} (#2) #3 }
\def\ijmpa#1#2#3{{\it Int. J. Mod. Phys.} {\bf A#1} (#2) #3 }

\def\bb#1{{\tt hep-th/#1}}


\def\ket{\rangle}
\def\bra{\langle}


\lref\rcon{A. Strominger, \npb {451}{1995}{109}, \bb{9504090.}}  
\lref\rgrossm{D.J. Gross and P.F. Mende, \plb {197}{1987}{129\semi}
 \npb {303}{1988}{407.}}
\lref\rgrossmanes{D.J. Gross and J.L. Ma{\~n}es, \npb {326}{1989}{73.}}
\lref\rshen{S.H. Shenker, {\it`` Another Length Scale in String
Theory?"}, Rutgers preprint RU-95-53, \bb{9509132.}}
\lref\rigor{I.R. Klebanov and L. Thorlacius, {\it ``The Size of
p-Branes"}, Princeton preprint PUPT-1574, \bb{9510200\semi}
S.S. Gubser, A. Hashimoto, I.R. Klebanov and J.M. Maldacena, {\it
``Gravitational lensing by p-Branes"}, Princeton preprint PUPT-1586,
\bb{9601057.}}
\lref\rbachas {C. Bachas, {\it ``D-Brane dynamics"}, ITP and Ecole
Polytechnique preprint, NSF-ITP/95-144, CPTH-S388.1195, 
  \bb{9511043\semi
} T. Banks and L. Susskind, {\it ``Brane-Antibrane Forces"}, Rutgers
preprint RU-95-87, \bb{9511194.}}
\lref\rdbrane{J. Dai, R.G. Leigh and J. Polchinski, \mpla
{4}{1989}{2073\semi}
R.G. Leigh, \mpla {4}{1989}{2767.}}
\lref\rpolch{J. Polchinski, {\it ``Dirichlet-Branes and Ramond-Ramond
Charges"}, NSF Institute for Theoretical Physics preprint NSF-ITP
95-122,
\bb{9510017.}}
\lref\rvene{D. Amati, M. Ciafaloni and G. Veneziano, \plb
{197}{1987}{81\semi}
\ijmpa {3}{1988}{1615.}}
\lref\rmende{P.F. Mende, \plb {326}{1994}{212.}}
\lref\rgreen{E. Corrigan and D. Fairlie, \npb {91}{1975}{527\semi}
M.B. Green,  \plb {69}{1977}{89\semi} 
M.B. Green, \plb {266}{1991}{325.}}


\line{\hfill PUPT-96-1590}
\line{\hfill {\tt hep-th/9601098}}
\vskip 1cm

\Title{\vbox{\baselineskip 12pt\hbox{}
 }}
{\vbox {\centerline{D-brane form factors }
\medskip
\centerline{at high energy }  }}

\centerline{$\quad$ {\caps J. L. F. Barb\'on}}
\smallskip
\centerline{{\sl Joseph Henry Laboratories}}
\centerline{{\sl Princeton University}}
\centerline{{\sl Princeton, NJ 08544, U.S.A.}}
\centerline{{\tt barbon@puhep1.princeton.edu}}
\vskip 0.4in

We study the high energy, fixed angle, 
 asymptotics of D-brane form factors to all
orders in string perturbation theory, using 
the Gross-Mende saddle point techniques. 
 The effective interaction size of all 
 D-branes grows linearly with the energy as  
 $\alpha' E/n$, where $n$ is the order of perturbation
theory, 
 except for the D-instanton, whose form factor is dominated by
end-point contributions, and remains point-like at high orders. The
qualitative features are independent of the R-R or NS-NS character of
the states used to probe the D-brane.


\Date{1/96}



An interesting question about D-branes, the stringy solitons carrying
R-R charge in type II string theories \refs\rdbrane \refs\rpolch, is their
physical size. In particular, it was observed by Shenker in
 \refs\rshen\  
that the physics of the R-R condensates \refs\rcon\ 
could be a hint at a new
dynamical scale in string theory, of order $\lambda \sqrt{\alpha'}$,
where $\lambda$ is the string coupling. 
 In a technical sense D-branes are infinitely thin, because
they are introduced as a sharp Dirichlet boundary condition for string
propagation. This is analogous to the fact  that an elementary string
has zero thickness, by contrast to  regular
solitonic solutions of
 the low energy effective field theory. In practice,
 interactions set the physical size of elementary strings to be
of
order $\sqrt{\alpha'}$. Likewise, since D-branes interact via string
exchange, we expect their physical size to be of order
$\sqrt{\alpha'}$ as well \refs\rbachas. As a direct check of this one
could calculate the form factor of a static D-brane, as probed by
string scattering \refs\rigor. Interestingly enough, the leading form
factor   shows
stringy features for all D-branes except the D-instanton, which
exhibits point-like behavior \refs\rgreen.
 This is surprising in the light of the
heuristic comments above. 

A quick way of searching for stringy form factors is to identify soft
high energy behavior.
String scattering amplitudes at high center
 of mass squared energy $s$ and fixed
angle $\phi$ have the form
\eqn\exp
{{\cal A}_g \sim e^{-s \alpha'  f(\phi)/ (g+1)  }
}
where $ g+1$ is
the order   in   
perturbation theory. This result was derived by Gross and Mende
\refs\rgrossm\
 by means of a saddle point evaluation of the worldsheet
path integral (see also \refs\rvene).
 The existence of a saddle point in the $interior$ of
the moduli space is a signature of stringy behavior. On the other
hand, we would interpret  a power fall off with energy  
as a signature of a point-like
object, by contrast to the soft high energy
 dependence displayed in \exp.  

It is very easy to adapt these techniques to the study of the D-brane
form factor at high energy and fixed angle. An interesting aspect of this
regime is its universality. If a Gross-Mende
 saddle exists in the interior of moduli
space, then it is universal in the sense
 that it is independent of the particular
string theory we are considering
 (one can work with the bosonic theory), and
furthermore it is independent of the states in external legs. 
The exponential term   \exp\  
only depends on the tachyonic part of the vertex operator,
 and so the results 
apply to R-R external particles as well as NS-NS.

We consider the two point function of string states with the
appropriate D-brane boundary conditions on $X^{\mu}$: Neumann for $\mu
= 0,...,p$; Dirichlet for $\mu = p+1,...,d-1$.
In the s channel this will be interpreted as the scattering of a
string state of momentum $p_1$ off a heavy stationary p-brane which
absorbs $q=-p_2 - p_1$ momentum. Split the d-dimensional vectors $p_i$
into Neumann and Dirichlet components as
\eqn\split
{p_i = (p^0_i, {\vec N}_i, {\vec D}_i) }
The Neumann components are conserved $p^0_1 + p^0_2 =0= {\vec N}_1 +
{\vec N}_2 $ and the p-brane absorbs the Dirichlet components of
momentum $q = (0,0,-{\vec D}_1 - {\vec D}_2)$. Define the momentum
transfer squared
\eqn\te
{t= -q^2 = -{\vec D}_1^2 -{\vec D}_2^2 -2{\vec D}_1 \cdot {\vec D}_2 }
and the ``transverse" invariant energy squared
\eqn\ese
{s= - (p_1^0)^2 + {\vec N}_1^2 = -(p_2^0)^2 + {\vec N}_2^2 }
These quantities satisfy ${\vec D}_i^2 = s-m_i^2$ and $2{\vec D}_1
{\vec D}_2 = -t-2s+m_1^2 + m_2^2$. 
We will consider ``frontal" scattering ${\vec N}_i =0$ for which the
following relations hold at high energy:
\eqn\rels
{{\vec D}_1 \cdot {\vec D}_2  \sim  -s\, {\rm cos}\,\phi\,\,\,,\,\,\,\,
\, 
t \sim -4\,s\, {\rm sin}^2\, {\phi\over 2}}
where $\phi$ is the scattering angle between the ingoing and outgoing
strings, to be held fixed in the high energy limit.  
This is the situation in which all the products $p_i \cdot p_j$ become
large as compared to the masses $m_i^2 = -p_i^2$,  and the world sheet
path integral can be dominated by a saddle point. The general 
amplitude has the form 
\eqn\mod
{{\cal A}_{\chi}
 = \lambda^{2g +B+C+N-2} \int d\mu\, \prod_{i,j}^{N} \, 
 e^{-{1\over 2} p_{i}^{\mu}
p_{j}^{\nu} \bra X_{\mu}(z_i) X_{\nu} (z_j)\ket}}
where $d\mu$ is some measure in the moduli space of Riemann surfaces
with $g$ handles, $B$ boundaries, $C$ cross-caps and $N$ punctures
where the vertex operators are inserted\foot{In the following, we will
focus on orientable world sheets, as appropriate for type II
strings.}. This measure has polynomial 
dependence on the  external momenta and polarizations.
 The saddle point 
is thus given by the (unstable) equilibrium configuration of  
$N$ Minkowskian charges with Coulomb interactions. 
For closed strings, these 
 saddles where identified by Gross and Mende as the n-fold 
branched covers
of the sphere with a  ${\bf Z}_{n}$ automorphism:
\eqn\cover
{y^{n} = \prod_{i=1}^{L} (z-a_i )^{L_i} }
where $\sum_i L_i = 0\, {\rm mod}\,( n)$,  all $L_i$ are coprime with
$n$, and the operator insertions lie at the branch points.
In general, the genus of such a curve is given by the Riemann-Hurwitz
formula  $g= 1-n+B/2$ where $B$ is the total branching number,
counting multiplicity. For the case above $(L_i,n)=1$ and all branch
points have maximum order $n-1$, leading to $g= (n-1)(L/2 -1)$. In
particular, with four branch points we have $g=n-1$. 
The solution to the electrostatic problem is then trivial since
the charges are located at the positions of the branch points. A
gaussian curve has length proportional to $g+1$ (it must go over each
sheet before it closes), so the electric fields scale like $1\over
g+1$. The area of the surface is  proportional to $g+1$, 
 and the 
corresponding electrostatic energy is  given by
\eqn\energy
{{\cal E}_{g} = {1\over 2} \sum_{i, j} p_i^{\mu} \cdot p_j^{\nu} \bra
X_{\mu} (z_i) X_{\nu} (z_j)\ket_g = { {\cal E}_0 \over g+1}}   
Branch points without associated charges do not contribute to the energy
but increase the genus of the surfaces so that, at a given order in
perturbation theory,  they are  
subdominant with respect to the $L=N   $ case.  
In this way one can easily understand the higher genus saddles in terms
of the sphere amplitude. It is also possible to deal with open string
boundaries and cross-caps \refs\rgrossmanes\ 
 by  simply using a generalization of the Schwarz reflection
principle. The open string saddles are obtained from 
 closed string saddles which are reflection symmetric about the real
axis. In the electrostatic problem, one is effectively using the method
of images to calculate the potential energy in the presence of a
set of conducting boundaries (along the real axis). So, for each open
string saddle world sheet, there is a closed string saddle obtained by
doubling the previous one. The generalization
from Neumann
 to Dirichlet boundaries is straightforward from this point of view,
because they simply correspond to perfectly  insulating
  boundaries\foot{As
expected, under T-duality Dirichlet and Neumann conditions are
interchanged, as well as momentum and winding modes in external states.
So we have an electric-magnetic duality transformation on the
world sheet \refs\rmende.}.  
We may consider closed string saddles with four branch points
located along the imaginary axis, such that the branched cover is
symmetric with respect to reflection on the real axis. By an
appropriate relabeling of the sheets we can always choose $L_1 =1$ in
\cover. Thus, we consider curves of the  form: 
\eqn\curve
{y^{g_0 +1} = \left({z-ia_1 \over z+ia_1}\right) \left({z-ia_2 \over
z+ia_2}\right)^{L_2} }
In general, if  we cut a genus $g_0$ closed orientable surface
  through the reflection plane we
obtain a surface with $g_2 +1$ boundaries and $g_1$ handles, such that $g_0
= 2g_1 + g_2$. The Euler number of the open Riemann surface is $\chi =
2-2g_1 -(g_2 +1) = 1-g_0$, and we find $2-\chi = 1+g_0$. 
Using \energy\ we conclude that the exponential term at the saddle point
is given by 
\eqn\hh
{e^{-{\cal E}_{\chi}} = e^{- {{\cal E}_0 \over 2-\chi}}}
In the case at hand, ${\cal E}_0$ is the electrostatic energy of the
upper half plane
  with two charges $p_1$ and $p_2$. The relevant Green function is
\eqn\green
{\bra X^{\mu} (z) X^{\nu} (w)\ket = -\alpha' \eta^{\mu\nu} \left( {\rm
log}|z-w| \pm {\rm log} |z-{\overline w}| \right)}
where the $\pm$ corresponds to the Neumann or Dirichlet components
respectively. The resulting electrostatic energy uses image charges
$p'_i = (p_i^{0}, {\vec N}_i , -{\vec D}_i)$ and has the form
\eqn\elen
{{\cal E}_0 = -{\alpha' \over 2} \sum_{i\neq j} p_i \cdot p_j\,
 {\rm log}|z_i
-z_j| - {\alpha' \over 2} \sum_{i,j} p_i \cdot p'_j \, {\rm log}|z_i -
{\overline z}_j|}
where singular self-energy terms have been subtracted. 
We can now use the $SL(2,{\bf R})$ symmetry to fix the second charge at
$z_2 = i$ and constrain $z_1 = iy$ to lie in the positive imaginary
axis\foot{The electrostatic problem has $SL(2,{\bf R})$ symmetry
precisely in the case of Minkowskian charges, where we can neglect
$p_i^2$ as compared to $p_i \cdot p_j$ \refs\rgrossm.}.
 In  the high energy limit $s\rightarrow \infty$ with
fixed angle $\phi$,  
\eqn\limit
{{{\cal E}_0\over 2\alpha'} 
 \rightarrow s\,{\rm log}\,2  +{t\over 4} \, 
 {\rm log}|y-1| - (s+t/4)\, 
{\rm log} |1+y| + {1\over 2}\, s\, {\rm log}\,y }
The  saddle point is located at ${\hat y} = e^{\pm i\phi}$,  
  outside the integration domain for
$y$, and  we need to distort the contour. The real part of the energy at
the saddle is
\eqn\ensad
{{\hat {\cal E}}_0 = -\alpha' s \left( {\rm sin}^2 {\phi \over 2}\, {\rm
log}\,{\rm sin}^2 {\phi\over 2} + {\rm cos}^2 {\phi\over 2}\,{\rm log}
\,{\rm
cos}^2 {\phi\over 2}\right) }
There is also a phase which combines with others  coming from the
fluctuation determinant around the saddle point,  producing the complete
phase
following 
from the Stirling approximation of the expression: 
$$
{\Gamma (-\alpha' s)\Gamma (-{\alpha' t\over 4}) 
\over \Gamma (-\alpha' s -
{\alpha' t\over 4})}
$$
which occurs in the exact evaluation of the disk form factor
 \refs\rigor.
Finally, putting everything together we arrive at the final result for
the leading exponential dependence of the form factor at order $2-\chi$
in perturbation theory:
\eqn\final
{{\cal A}_{\chi} \sim \lambda^{2-\chi}\, 
 {\rm exp} {\alpha' s\over 2-\chi}
\left({\rm sin}^2 {\phi \over 2}\,{\rm log}\,{\rm sin}^2 {\phi\over 2} +
{\rm cos}^2 {\phi\over 2}\, {\rm log}\,{\rm cos}^2 {\phi\over 2}\right) }
We see that the form factor is characterized by the string scale
$\sqrt{\alpha'}$ independently of the 
particular states (NS-NS  or R-R)  we use to probe the D-brane.
The genus dependence in \final\ 
is precisely the same that one finds in high energy
open string scattering \refs\rgrossmanes. This is hardly a surprise,
given the relation between open strings and D-branes via T-duality.

 It is interesting to study the structure of the classical saddle world
sheets:
\eqn\class
{X^{\mu}_{c\ell} (z) = {i\alpha' \over 2-\chi} \sum_{j=1}^N p^{\mu}_j  
\left({\rm log}(|z-a_j| \pm {\rm log}|z-{\overline a}_j|\right)
}
For the closed string scattering with world sheet $y^{g_0 +1} =
\prod_{i=1}^{4} (z-a_i)^{L_i}$,   two strings wound $g_0 +1$
times  interact in a one-string irreducible vertex and propagate
an intermediate state of several closed strings \refs\rgrossm.
The effective vertices in a particular channel $i+j$ can be obtained
from the degeneration in which the branch points $a_i$ and $a_j$
coincide. The branching number of the resulting branch point $a_{i+j}$
depends on whether $g_0 +1$ and $L_i + L_j$ have a common factor, and
it is given by $B_{i+j} = g_0 +1-(g_0 +1, L_i + L_j)$.
Applying the Riemann-Hurwitz formula we end up with an effective vertex
of genus $g_{i+j} = B_{i+j} /2$, and $g_0 -2g_{i+j} + 1= (g_0 +1, L_i
+ L_j)$ intermediate
propagating strings. For the curves above \curve, we get genus zero
effective vertices in the $s$ channel, with  $M =
(2-\chi, 1+L_2) $
intermediate open strings propagating along the D-brane, and $
M' = (2-\chi 
-M)/2$ closed strings exchanged.  
 In  the $t$ channel we have a genus $M'$  effective vertex,  
and $M$ closed strings exchanged with the D-brane.
 The important feature
from the physical point of view is that, according to the formula
\class, the classical world sheet saturating the form factor has a
scale $\bra X\ket \sim ({\rm energy})/(2-\chi)$, whereas the Schwarzschild
radius for that energy scales like $\lambda^2 ({\rm energy})$. So, no matter
how small we take the string coupling, there is a potential breakdown
of perturbation theory at order $2-\chi \sim \lambda^{-2}$, due to
strong gravity effects.

We may also  compare formula \final\ with the Regge
approximation. In the limit of small scattering angles $-t << s$ it is
reasonable to regard the higher genus amplitude as a multi-scattering
process where $2-\chi$ closed strings are exchanged with the
D-brane. In fact, this is the structure of the dominant saddle world
sheets without effective vertices in the $t$ channel. To lowest order
$$
{\cal A}_0 \sim \lambda \, s^{\alpha' t /4}
$$
We then estimate the amplitude as
$$
{\cal A}_{\chi} \sim {\rm Max}\, \lambda^{2-\chi} \, s^{(\alpha' \sum_i
t_i)/4}
$$
subject to the constraint $\sum_{i=1}^{2-\chi} \sqrt{-t_i} \leq
\sqrt{-t}$. The solution is $t_i = t/(2-\chi)^2$ and we find 
$$
{\cal A}_{\chi} \sim \lambda^{2-\chi}\, s^{\alpha' t \over 4(2-\chi)}
$$
in perfect agreement with \final. It is amusing to notice that the
effective Regge slope for the interaction with the D-brane is
$\alpha'_D = \alpha'/2$, precisely the T-dual of the open string Regge
slope $\alpha'_{\rm open} = 2\alpha' = \alpha'^2 /\alpha'_D$.

One could easily generalize these results to non-orientable
world-sheets (along the lines of ref. \refs\rgrossmanes) with similar
qualitative behavior. Also, more general amplitudes can be considered.
For example, ``deep inelastic" processes in which we have many strings
in the final state. Unfortunately, as for the case of multi-string
scattering, the corresponding saddle points are difficult to
characterize analytically.
One could also discuss amplitudes with Dirichlet boundaries located at
different points in the target space (several D-branes). 
However, such calculations have a limited physical interest, because
they do not really capture D-brane dynamics. The saddle point
techniques regard the D-branes as static objects capable of absorbing
any amount of energy-momentum. This picture is precisely wrong in the
high energy limit. Here, we have used the high energy limit only as a
device     to extract
some general features of the form factors at high orders in
perturbation theory.

The methods used imply that the D-instanton remains point-like (power
behaved form factor at high energy) to all orders in perturbation
theory. In this case we have Dirichlet boundary conditions in all
coordinates and the high energy limit of the electrostatic energy
degenerates to
$$
{\cal E}_{\chi} \rightarrow -{\alpha' t\over 2(2-\chi)}
\, {\rm log}{\left|1+y\over 1-y\right|}
$$
The  integral over $y$ is dominated by the end-point contribution at
$y=0$ and the saddle point at $y=\infty$, both of zero energy, which
explains the power-like behavior of the amplitude. In the fully
Dirichlet case, the image charge is completely inverted $p' = -p$ and it
is thus natural that no equilibrium configuration of separate charges
exists: the vertex operators tend to collapse on  the Dirichlet
boundary. For p-branes with $p>-1$, some of the components of the image
charges have the same sign, and there is a compromise between repulsion
and attraction leading to a non-trivial saddle point in the interior of
moduli space.

\newsec{Acknowledgements}
It is a pleasure to thank D. Gross,   I. Klebanov and S. Ramgoolam
 for useful
discussions. This work was supported by NSF PHY90-21984 grant.
 
\listrefs
\bye